\documentclass[twocolumn,times,tighten]{aastex631}
\usepackage{color}
\usepackage{enumitem}

\usepackage{hyperref}
\usepackage{verbatim}
\usepackage{amsmath,amstext,mathrsfs}
\usepackage[all]{hypcap} 
\usepackage{afterpage}    
\usepackage{marginnote}
\usepackage{makecell}
\usepackage{subfigure}
\usepackage{xfrac}
\defcitealias{LP00}{LP00}
\defcitealias{LP12}{LP12}
\defcitealias{GS95}{GS95}
\defcitealias{LV99}{LV99}
\defcitealias{KLP16}{KLP16}
\defcitealias{KLP17a}{KLP17}
\defcitealias{PaperI}{Paper I}
\DeclareFontFamily{OMS}{oasy}{\skewchar\font48 }
\DeclareFontShape{OMS}{oasy}{m}{n}{%
         <-5.5> oasy5     <5.5-6.5> oasy6
      <6.5-7.5> oasy7     <7.5-8.5> oasy8
      <8.5-9.5> oasy9     <9.5->  oasy10
      }{}
\DeclareFontShape{OMS}{oasy}{b}{n}{%
       <-6> oabsy5
      <6-8> oabsy7
      <8->  oabsy10
      }{}
\DeclareSymbolFont{oasy}{OMS}{oasy}{m}{n}
\SetSymbolFont{oasy}{bold}{OMS}{oasy}{b}{n}

\DeclareMathSymbol{\smallleftarrow}     {\mathrel}{oasy}{"20}
\DeclareMathSymbol{\smallrightarrow}    {\mathrel}{oasy}{"21}
\DeclareMathSymbol{\smallleftrightarrow}{\mathrel}{oasy}{"24}

\newcommand*{\kh}{} % This is KH's color

\newcommand*{\yh}{\bf \color{purple} Yue:}
\newcommand*{\al}{\bf \color{red} AL:}
\newcommand*{\DP}{\bf \color{olive} DP:}

\newcommand*{\torefereeone}{\color{red} \bf}

\graphicspath{{./}{Fig/}}

\shorttitle{Measuring magnetic field using gradients}
\shortauthors{Lazarian, Hu \& Pogosyan}
\begin{document}

\title{Obtaining Magnetization of Super-Alfv\'enic Turbulence with the Structure Functions of Gradient Directions}

\author{A. Lazarian}
\affiliation{Department of Astronomy, University of Wisconsin-Madison, USA}

\author[0000-0002-8455-0805]{Yue Hu*}
\affiliation{Institute for Advanced Study, 1 Einstein Drive, Princeton, NJ 08540, USA }

\author{D. Pogosyan}
\affiliation{Department of Physics, University of Alberta, Canada}

\email{lazarian@astro.wisc.edu, yuehu@ias.edu, pogosyan@ualberta.ca; *NASA Hubble Fellow}

\begin{abstract}
Super-Alfv\'enic turbulence is widespread in astrophysical environments, including molecular clouds and the diffuse plasma of galaxy clusters. At large scales, magnetic fields play only a minor dynamical role; however, for sufficiently extended turbulent cascades, the motions transition into the MHD regime at a characteristic scale $l_A$. We introduce a new diagnostic based on the structure functions of the gradient directions, which can be obtained directly from spectroscopic and synchrotron intensity observations. We demonstrate that the new measure robustly recovers the transition scale $l_A$. Building on this result, we propose a generalized expression that replaces the traditional Davis–Chandrasekhar–Fermi (DCF) method for estimating magnetic field strength in the super-Alfv\'enic regime, where the DCF approach fails. We further show how the magnetization and magnetic field strength of diffuse media, such as the intracluster medium, can be inferred using synchrotron intensity maps. Our theoretical predictions are validated through numerical simulations.
\end{abstract}

\keywords{Interstellar medium (847) --- Interstellar magnetic fields (845) --- Plasma astrophysics (1261) --- Magnetohydrodynamics (1964)}

\section{Introduction}
\label{sec:intro}
Astrophysical magnetic fields are ubiquitous, and their influence on cosmic environments is profound. Magnetic fields regulate multiple stages of star formation (e.g., \citealt{1956MNRAS.116..503M,2006ApJ...647..374G,2006ApJ...646.1043M,2007IAUS..243...31J}) and play a central role in structuring the interstellar and intracluster media. Because astrophysical plasmas typically have extremely large Reynolds numbers, magnetized flows are generically turbulent (see \citealt{2004ARA&A..42..211E,MO07,2021NatAs...5..342M,2022ApJ...941..133H,2025ApJ...994..193B,2025ApJ...986...62H}). Turbulence itself is well established observationally through measurements of density fluctuations (e.g., \citealt{1995ApJ...443..209A,CL09}), velocity statistics (e.g., \citealt{1981MNRAS.194..809L,2004ApJ...615L..45H,2010ApJ...710..853C,VDA,cattail,spectrum}), and synchrotron emission (\citealt{2013A&A...558A..72I,LY18polar,2022ApJ...940..158W}).

A key parameter characterizing magnetized turbulence is the Alfv\'en Mach number, $M_A=V_L/V_A$, where $V_L$ is the injection velocity and $V_A$ is the Alfv\'en speed (see \citealt{BeresnyakLazarian+2019}). The value of $M_A$ quantifies the dynamical importance of magnetic fields. When $M_A<1$, turbulence is sub-Alfv\'enic and motions are magnetically dominated at all scales. In contrast, when $M_A>1$, turbulence is super-Alfvénic: magnetic fields influence the dynamics only below the transition scale $l_A=LM_A^{-3}$, where turbulent motions become magnetically coupled. In both regimes, magnetic fields remain critical for key astrophysical processes such as star formation, cosmic-ray transport, and the diffusion of matter and heat. Accurately determining $M_A$ from observations is therefore essential, yet remains a major challenge. Developing robust diagnostics of magnetization is central to advancing our understanding of turbulent astrophysical environments.

Obtaining the magnetic field strength from observations is a separate challenge. 
The technique that uses fluctuations in polarization angles together with fluctuations in Doppler-shifted velocities was first proposed by \citet{1951PhRv...81..890D} and independently by \citet{CF53} to estimate magnetic fields in molecular clouds. This method, now known as the Davis--Chandrasekhar--Fermi (DCF) technique, relies on the assumption that magnetic-field and velocity fluctuations are related through the Alfv\'en relation \citep{1942Natur.150..405A}:
\begin{equation}
\delta B \;=\; \delta v \, \sqrt{4\pi \langle\rho\rangle},
\label{eq:alf}
\end{equation}
where $\langle\rho\rangle$ is the mean mass density. This relation is valid only in the sub-Alfv\'enic regime ($M_A < 1$). In addition, the DCF method requires both polarimetric and spectroscopic observations of the same region, limiting its applicability in many astrophysical environments.

Estimating the magnetization of super-Alfv\'enic turbulence ($M_A > 1$) remains particularly difficult \citep{LPY25}, even though turbulence in molecular clouds is often super-Alfv\'enic \citep{padoan16_turb}, and turbulence in galaxy clusters is also known to be super-Alfv\'enic (see \citealt{Brun_Laz07}). The DCF method fails in this regime for the reasons noted above, underscoring the need for new techniques capable of probing magnetic fields and magnetization in $M_A>1$ environments.

In this paper, we investigate whether the structure functions of gradient directions can be used to retrieve $M_A$. We apply both velocity gradients  \citep{YL17a, LY18polar, 2018MNRAS.480.1333H,Hu_mapping23} and synchrotron-intensity gradients \citep{synch_grad17, Hu_clusters24} as tracers. While gradients trace magnetic-field morphology, their statistical behavior is not identical to that of the magnetic field itself, as shown in \citet{Laz_Y_P25}. We leverage these differences to infer both the Alfv\'en Mach number and the magnetic-field strength from the structure functions of gradient directions.

In \S 2, we summarize the properties of superAlfvenic turbulence and the use of gradients for magnetic field tracing. In \S 3 we introduce the new measures, the structure function, and the spectrum of gradient directions.

\section{Properties of super-Alfvenic MHD turbulence: the transitional scale $l_A$}
Within the modern understanding of MHD turbulence \citep{BL19}, the turbulent cascade is a superposition of three interacting cascades corresponding to the Alfv\'enic, slow, and fast modes. Because strong nonlinear interactions cause Alfv\'enic perturbations to decay within approximately one wave period, the Alfv\'enic cascade dominates many key physical processes. The back-reaction of the slow and fast modes on the Alfv\'enic cascade is marginal in the regime of strong Alfv\'enic turbulence (\citetalias{GS95}; \citealt{2002ApJ...564..291C,leakage}; see also the appendix of \citealt{2024arXiv240517985P}). Therefore, when discussing super-Alfv\'enic turbulence, we primarily focus on the properties of the Alfv\'enic cascade.

Super-Alfv\'enic turbulence can be viewed as consisting of two sequential cascades: a hydrodynamic cascade at large scales and an MHD cascade at smaller scales. The transition between these regimes occurs at the scale $l_A$, where the turbulent velocity becomes equal to the Alfv\'en speed. Using the Kolmogorov scaling of hydrodynamic turbulence,
\begin{equation}
    v_l \approx V_L \left( \frac{l}{L} \right)^{1/3},
    \label{v_GS}
\end{equation}
one finds that $v_l = V_A$ at
\begin{equation}
    l_A \approx L M_A^{-3}.
    \label{l_A}
\end{equation}

The above arguments assume that $l_A$ exceeds the dissipation scale $l_{\rm diss}$, a condition we adopt throughout this paper. For isotropic driving, a turbulent volume of size $L^3$ contains approximately $M_A^9$ magnetic subdomains of size $l_A$. Within each $l_A$-domain, magnetic fields regulate fluid motions, while the relative orientations of the magnetic fields in neighboring domains are only weakly influenced by the large-scale mean field. This influence becomes negligible in the limit of large $M_A$.

In the \citetalias{GS95} picture of trans-Alfv\'enic turbulence, the wavevectors of Alfv\'enic fluctuations become increasingly perpendicular to the magnetic field as the cascade proceeds. These motions can be interpreted as eddies that shear and mix the magnetic field, as described in \citetalias{LV99}. The existence of such eddies follows naturally from the theory of turbulent reconnection (\citetalias{LV99}; see \citealt{2020PhPl...27a2305L} for a review), which predicts that magnetic reconnection proceeds on the timescale of a single eddy turnover. As a result, the magnetic field does not inhibit the eddies that mix plasma and magnetic flux perpendicular to the field lines surrounding each eddy.

This physical picture defines the {\it local} magnetic-field direction and the corresponding {\it local magnetic reference frame}, a key refinement of the original GS95 formulation, which treated fluctuations with respect to the global mean magnetic field.\footnote{The critical-balance condition relating parallel and perpendicular scales is not satisfied when measured in the global-mean-field frame.} The concept of a local magnetic field has been confirmed numerically by \citet{CV00} and subsequent studies \citep[e.g.,][]{MG01,2002PhRvL..88x5001C}, and is now an essential component of the modern understanding of the Alfv\'enic cascade. Because anisotropic Alfv\'enic eddies align with the local magnetic field, measurements of their anisotropy trace the detailed structure of the magnetic field rather than the large-scale mean field.

\section{Structure function and spectrum of the gradient directions}

\subsection{Velocity and synchrotron intensity gradients}

MHD turbulence predicts that in turbulent magnetized eddies, matter mixes perpendicular to the local magnetic-field direction. Such mixing naturally generates velocity gradients—and other types of gradients that are also oriented perpendicular to the magnetic field. This fundamental property underlies Gradient Theory (GT), which exploits the orientation of these gradients to trace the magnetic-field structure in observational data \citep{LYP24}.

In this paper, we discuss the most common types of gradients, the velocity gradients (VG; \citealt{YL17a, LY18a,2018MNRAS.480.1333H}) and synchrotron intensity gradients (SIGs; \citealt{synch_grad17,2024NatCo..15.1006H}). The latter represent the gradients of the magnetic field. For Alfv\'enic motions, the properties of magnetic fields and velocities are similar. Thus, both VGs and SIGs trace magnetic fields in a similar way. As we show further, this similarity in magnetic field tracing is sustained in the case of superAlfvenic turbulence.

\subsection{Gradients in super-Alfv\'enic turbulence}
\label{sec:supgrad}

Magnetic turbulence is inherently anisotropic, with turbulent eddies elongated along the magnetic-field direction. As a result, fluctuations in both velocity and magnetic field are stronger in the direction perpendicular to the field. This property forms the basis of the Gradient Technique (GT), which enables magnetic-field tracing using a variety of observational diagnostics \citep{synch_grad17,LY18a,LY18polar,2018MNRAS.480.1333H,2019ApJ...886...17H}. 
For simplicity, in what follows, we focus on gradients of velocity amplitudes and synchrotron intensities. These gradients are obtained through statistical averaging over sub-blocks of observational or synthetic data \citep{YL17b}. Previous work has shown that the probability distribution functions (PDFs) of gradients within such sub-blocks can be used to infer the Alfv\'en Mach number $M_A$ and thus the magnetization of the medium \citep{dispersion}. However, the sensitivity of gradient-PDF statistics to $M_A$ decreases in the super-Alfv\'enic regime ($M_A>1$).

In hydrodynamic turbulence with Kolmogorov scaling, velocity-amplitude gradients behave as 
\begin{equation}
GR_{3D-point} \sim v_l/l \sim (V_{inj}/L_{inj}) (L_{inj}/l)^{2/3} \sim l^{-2/3},
\end{equation}
where $L_{inj}$ and $V_{inj}$ are the injection scale and the injection velocity, respectively. Obviously, the value of the gradient is increasing towards smaller scales, i.e. $\sim l^{-2/3}$. The smallest eddies in {\it hydrodynamic} turbulence, the gradients of velocities map directions whose directions are isotropically distributed, such gradients add up in a random walk fashion:
\begin{equation}
    GR_{line-of-sight}\sim (V_{inj}/L_{inj}) (L_{inj}/l)^{7/6}
    \label{hydro1}
\end{equation}
which also increases as the eddy size decreases. 

\begin{comment}
However, in the Gradient Technique approach, sub-block averaging is employed \cite{YL17a}. The addition of gradients acts within a rectangular sub-block of the size $l_{sub}$x$  l_{sub}$ is a random walk addition, i.e.
\begin{equation}
    GR_{subblock}\sim (V_{inj}/L_{inj}) (L_{inj}/l)^{5/6}(L_{subblock}/l),
\end{equation}
which increases as $l$ approaches the sub-block size. For $l=L_{subblock}$, the contributions are added in a regular way, making the contributions eddies of sub-block size dominate the smaller eddies according to the ratio $(L_{subblock}/l)^{1/6}$. As a result, the gradient measurements correspond to the sub-block scale. 
\end{comment}

The hydrodynamic picture above also holds for magnetic fields that are passively advected by velocities on scales larger than $l_A$. In this case, the minimal hydrodynamic-like case is $l_A$, i.e. $l=l_A$ in Eq. (\ref{hydro1}). The velocities in the range of scales $[l; l_A]$ are coherent due to the dynamically important magnetic field in the $l_A$-domain. Thus, the contributions over $l_A$ is summed up linearly, i.e.
\begin{equation}
    GR_{domain} \sim (V_{inj}/L_{inj}) (L_{inj}/l)^{2/3} (l_A/l) (L_{inj}/l_A)^{1/2},
\end{equation}
which is a factor $(l_A/l)^{1/2}$ larger than the result that would be summation over random contributions from $l$-scale uncorrelated hydrodynamical eddies. 

%The gradient contributions from $l_A$-domains are added in a random walk direction, which makes the pointwise contributions from velocity gradients at scale $l<l_A$ a factor $(l_A/l)^{10/6}$ larger than the contribution of gradients from hydrodynamic eddies from scales larger than $l_A$. This allows sampling of magnetic field directions with velocity gradients that in $l_A$-domains are aligned perpendicular to the magnetic field on scales $l$. 

In the case of magnetic gradients that can be sampled by SIGs, the magnetic field gets a flat power spectrum at scales larger than $l_A$. The corresponding scaling corresponds to $b_A (l_A/l)^{1/2}$, where $b_A\sim V_{inj} M_A^{-1}$. This means that the contribution from the gradient arising from magnetic eddies larger than $l_A$ is further suppressed compared to the velocity gradients. As a result, SIGs, similar to velocity gradients, are dominated by gradients from scales less than $l_A$ that reflect the magnetic field structure.

\subsection{Structure function of gradient directions}

In this paper, we use gradients to construct a new statistical measure of magnetized turbulence: the Structure Function of Gradient Directions (SFGD),
\begin{equation}
D^\phi(\mathbf{R}) = \left\langle \sin^2(\phi_1 - \phi_2) \right\rangle,
\label{D}
\end{equation}
where $\phi_1 = \phi({\bf X})$ and $\phi_2 = \phi({\bf X}+{\bf R})$ denote the gradient directions at two positions separated by a plane-of-sky (POS) displacement ${\bf R}$. For uncorrelated gradient directions, $D^\phi$ saturates to the asymptotic value of $1/2$. This measure has previously been applied to characterize the statistics of polarization directions \citep{Laz_Pog22, Laz_Y_P25}.

\begin{comment}
Taking the Fourier transform of $D^\phi$ yields ${\cal F_\phi}$, which is the sum of the Fourier transforms of the correlation functions $\langle \cos 2\phi_1 \cos 2\phi_2 \rangle$ and $\langle \sin 2\phi_1 \sin 2\phi_2 \rangle$. These can be expressed using the pseudo-Stokes parameters $Q$ and $U$ defined for gradients in \citet{LY18a}:
\begin{equation}
\cos(2\phi) = \frac{Q}{\sqrt{Q^2 + U^2}}, \qquad
\sin(2\phi) = \frac{U}{\sqrt{Q^2 + U^2}}.
\label{eq:stokes1}
\end{equation}
As a result,
\begin{equation}
{\cal F_\phi}(K)
    = \left| F\left\{ \frac{Q + iU}{\sqrt{Q^2 + U^2}} \right\} \right|^2,
\label{eq:Fourier}
\end{equation}
where $F$ denotes the Fourier transform and $K$ is the magnitude of the POS wavevector.
\end{comment}

In this work, we examine the properties of $D^\phi(R)$, focusing on how gradient-direction correlations evolve with increasing line-of-sight separation.  

%When $R > l_A$, the lines of sight probe independent $l_A$-domains, and their gradient directions become uncorrelated. Consequently, $D^\phi(R)$ saturates to the value of $1/2$ for $R > l_A$. The turnover of $D^\phi(R)$ therefore provides a direct way to observationally determine the transition scale $l_A$.

Although gradients trace the magnetic-field orientation, their statistical properties differ from those of the magnetic field itself \citep{Laz_Y_P25}. We exploit this distinction to develop a new and robust method to infer $l_A$ and the Alfv\'en Mach number $M_A$ in super-Alfvénic turbulence. We further demonstrate how these tools can be used to estimate magnetic-field strengths in super-Alfvénic astrophysical environments.

%For instance, gradients of synchrotron intensity have proven to be a powerful tool for resolving the magnetic field structure of galaxy clusters (\cite{Hu_clusters24}). Therefore, studying the turbulence spectrum and the turbulence injection scale using PA fluctuations measured with synchrotron gradients is a promising way of probing turbulence on the largest astrophysical scales.

Our paper describes how the statistics of $D^{\phi}$ changes with the Alfven Mach number, $M_A=V_L/V_A$, where $V_L$ is the turbulent velocity at the injection scale $L$ and $V_A$ is the Alfven velocity.

%In \cite{Cho23} synthetic observations of MHD super-Alfvenic turbulence simulations were decomposed into the E and B modes used in cosmological CMB studies. \cite{Cho23} demonstrated numerically that the ratio of the amplitudes of these modes at small scales depended on the Alfven Mach number $M_A=V_L/V_A$, where $V_L$ is a turbulent injection speed and $V_A$ is Alfven speed. A theoretical understanding of this result is essential for obtaining media magnetization parameterized by $M_A$.

\begin{figure*}[ht]
\centering
\includegraphics[width=1.0\linewidth]{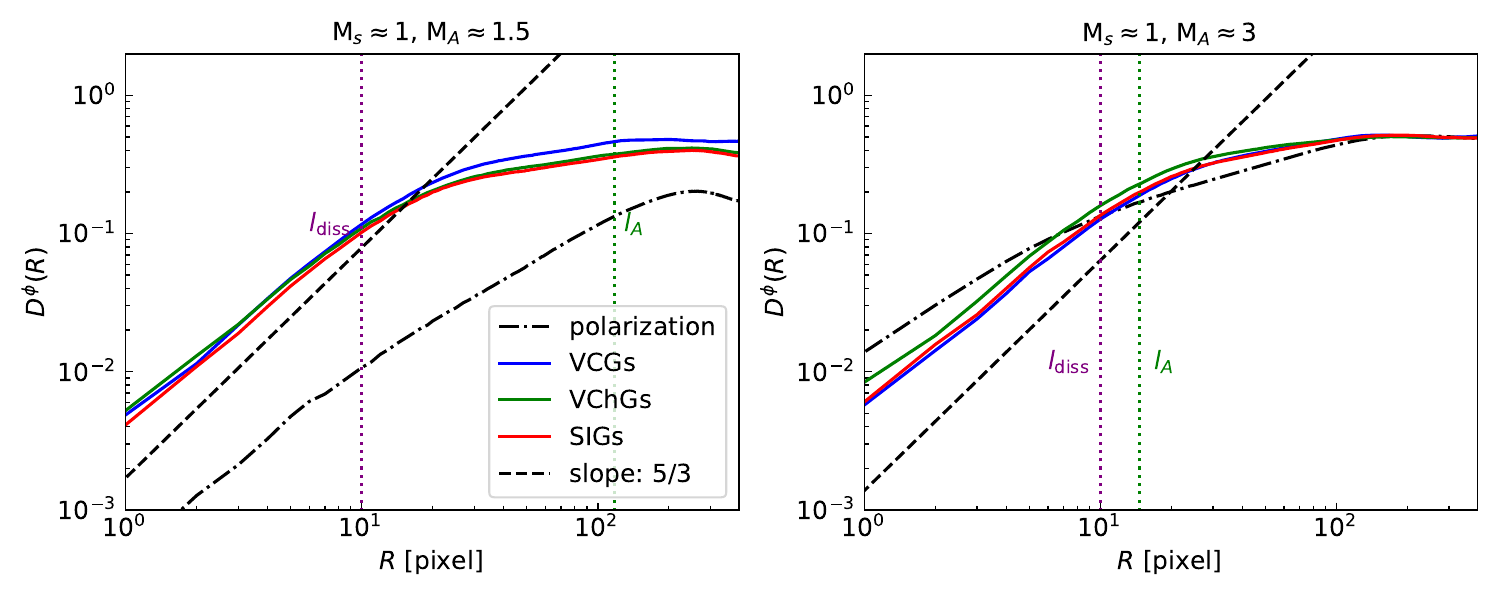}
\caption{Structure functions of velocity-gradient directions measured from velocity centroids and channel maps for $M_A=1.5$ and $M_A=3$. For comparison, we also show structure functions of magnetic-field directions (blue), which can be obtained observationally, and structure functions of velocity directions (red), which can only be measured from simulations.}
\label{fig:grad1}
\end{figure*}

\begin{comment}

\begin{figure*}[ht]
\centering
\includegraphics[width=1.0\linewidth]{fig_2024/Dspectrum.pdf}
\caption{\yh I did not see much difference between the two panels....}
\label{fig:grad1}
\end{figure*}
\end{comment}

\section{Numerical data sets}

We calculate gradients using the results of 3D superAlfvenic simulations described in detail \cite{LPY25} and presented in Table 1.  The results are obtained with AthenaK \citep{2024arXiv240916053S}, which solves the standard compressible isothermal MHD equations within a $792^3$ box with periodic boundaries. The turbulence driving is solenoidal with a peak wavenumber of $2 \times 2\pi/L_{\rm box}$.

\begin{table}
	\centering
	%\resizebox{\linewidth}{!}{
 \begin{tabular}{ | c | c | c | c | c |}
		\hline
		Run & $M_s$ & $M_A$ & $\beta$ & Resolution \\ \hline \hline 
		A1 & 1.0 & 1.5 & 4.5 & $792^3$ \\
		A2 & 1.0 & 3.0 & 18.0 & $792^3$ \\ 
        \hline
  	%A0 & 1.21 & 1.25 & 0.51 - 1.56 & 0.58 - 1.53 & AthenaK\\\hline
	\end{tabular}
 %}
	\caption{\label{tab:sim} Parameters of MHD turbulence simulations. The sonic and Alfv\'en Mach numbers are $M_s$ evaluated at a given snapshot, and $\beta = 2(M_A/M_s)^2$ is plasma magnetization.
 }
\end{table}
In the simulations, the sonic Mach number is $M_s \approx 1$ and the plasma beta is $\beta = 2(M_A/M_s)^2 > 1$, indicating that the gas pressure exceeds the magnetic pressure. The corresponding 3D velocity and magnetic-field spectra are presented in Fig.~1 of \citet{LPY25}. As theoretically expected, the velocity spectrum follows a Kolmogorov scaling throughout the inertial range, while the magnetic spectrum shows a clear break at the wavenumber $k \sim 1/l_A$.

We compute gradients using the sub-block averaging technique introduced in \citet{YL17a}. Its application to velocity-channel data is described in \citet{LY18a}, and its use for synchrotron intensities is presented in \citet{Hu_clusters24}. The sub-block averaging procedure reveals that, in a statistical sense, velocity and synchrotron-intensity gradients are preferentially oriented perpendicular to the magnetic field—an alignment that emerges from the anisotropic nature of MHD turbulence rather than from pointwise measurements.

\section{Results and analysis}

\subsection{Analysis of plots}

Fig.~\ref{fig:grad1} shows the structure functions of gradient directions measured from (a) velocity centroids, (b) velocity-channel, and (c) synchrotron intensity maps. The GT analysis was applied to data with $M_A = 1.5$ and $M_A = 3$. The vertical lines mark the scales where numerical dissipation becomes important ($l_{\rm diss}$) and where the turbulent velocity equals the Alfv\'en speed ($l_A$). 

The sturucure functions of gradient directions for velocity and synchrotron are similar, which corresponds to our expectations discussed in \S \ref{sec:supgrad}. The properties of sychrotron polarization directions are different, which corresponds to the different ways gradients and polarization trace magnetic fields.

As discussed in \citet{LPY25}, the structure functions below the dissipation scale follow an $R^2$ scaling—coincidentally very similar to the $R^{5/3}$ scaling expected for projected Kolmogorov turbulence.

Over the range $[l_{\rm diss},\,l_A]$, we find that structure functions of all gradient types exhibit spectra shallower than the Kolmogorov expectation. The structure functions saturate at $R\approx l_A$, indicating that correlations in gradient directions persist only up to the Alfv\'enic transition scale. This behavior provides a practical method to determine $l_A$ directly from observations.

\subsection{Obtaining turbulence magnetization}
If the scale $l_A = L M_A^{-3}$ is determined from the analysis of $D^\phi$, this immediately provides a way to infer the Alfv\'en Mach number:
\begin{equation}
    M_A \approx \left( \frac{L}{l_A} \right)^{1/3}.
    \label{eq:machA}
\end{equation}

For molecular clouds, the turbulence injection scale $L$ can be obtained from the structure functions of velocity centroids or density fluctuations. For galaxy clusters, $L$ can be inferred from the structure functions of synchrotron intensities or from the structure functions of polarization directions (see \citealt{LPY25}). 

The essence of this method is that $D^\phi(R)$ saturates at the Alfv\'enic transition scale $l_A$, whereas other structure functions saturate at the injection scale $L$. Measuring both saturation scales therefore enables a direct observational determination of $M_A$.

\subsection{Synergy with other approaches}
$D^\phi$ thus provides a new avenue for determining $M_A$. Interestingly, this method complements another approach for estimating $M_A$ based on the PDFs of gradient amplitudes, which has been shown to recover the magnetization of the medium from observational data \citep{dispersion,LYP24}. The availability of two independent techniques for measuring $M_A$ enhances the robustness and accuracy of magnetization estimates.

Conversely, if $M_A$ is obtained through an alternative method, the measurement of $l_A$ enables a new way to infer the turbulence injection scale $L$. More generally, because $l_A$ depends on both $L$ and $M_A$, its determination is synergistic with other observational diagnostics of these parameters, strengthening the overall toolkit for characterizing magnetized turbulence in astrophysical environments.

\section{Practical benefits: Obtaining magnetization and magnetic field strength}

\subsection{Measurements of POS and 3D magnetization}
.

The extent to which magnetic fields regulate turbulent motions is determined by the value of the Alfv\'en Mach number $M_A$. For $M_A > 1$, magnetic fields are tangled and chaotic on scales larger than $l_A$, which significantly influences a wide range of astrophysical processes, including the propagation of cosmic rays \citep{YL03,LX22}, heat transport \citep{Lazarian06}, reconnection-driven diffusion of magnetic fields \citep{Laz05,Laz14r,Sant10,San13}, and the damping of Alfv\'en waves along with the suppression of the streaming instability \citep{Lazarian16}. Thus, even when the magnetic-field strength is not directly known, the value of $M_A$ alone permits quantitative assessment of many astrophysically important properties of magnetized media.

The GT applied to velocity channel maps has been used to recover the three-dimensional magnetic-field structure by exploiting Galactic rotation in diffuse media \citep{Hu_mapping23}, and to map magnetic fields traced by different molecular species \citep{2019ApJ...873...16H,2022MNRAS.511..829H}. Three-dimensional magnetic-field distributions can also be obtained using gradients of synchrotron polarization \citep{LY18polar}. In all these cases, the structure function of gradient directions and its associated spectrum provide a unified tool for probing the spatial distribution of the transition scale $l_A$ and the magnetization parameter $M_A$ across astrophysical environments

\subsection{Extension of DCF-type approach}
The DCF technique \citep{Davis51,CF53} does not apply to super-Alfv\'enic turbulence, because its central assumption—the equipartition between kinetic and magnetic energies—is strongly violated at the injection scale. However, equipartition is restored at the transition scale $l_A$, where the turbulent velocity equals the Alfvén speed. At this scale, the velocity dispersion satisfies $\langle (\delta v)^2 \rangle \simeq V_A^2$, enabling an alternative estimate of the magnetic field strength using the structure function of velocity centroids,
\begin{equation}
    D^v({\bf R}) = \left\langle \big(V({\bf X}_1) - V({\bf X}_2)\big)^2 \right\rangle .
    \label{Dv}
\end{equation}

Once $l_A$ is determined, the magnetic-field strength can be obtained by evaluating the structure function at this scale,
\begin{equation}
    B \approx 2 \sqrt{\pi \rho\, D^v(l_A)},
    \label{B1}
\end{equation}
which mirrors the logic of the DCF method, but replaces the global velocity dispersion with the locally measured value $D^v(l_A)$.

A key advantage of Eq.~(\ref{B1}) is that, unlike Eq.~(\ref{eq:machA}) for determining $M_A$, it does not require prior knowledge of the turbulence injection scale $L$. This formulation therefore enables spatially resolved mapping of the magnetic-field strength with an effective resolution set by $l_A$.

\subsection{MM2 approach}
Instead of relying on the velocity dispersion, one may use the sonic Mach number $M_s$, which can also be obtained observationally. For example, by analyzing the statistics of synchrotron intensity fluctuations \citep{2011ApJ...736...60T,2011Natur.478..214G}, one can estimate $M_s$ and subsequently determine the magnetic-field strength using the MM2 approach introduced in \citet{Laz_Y_P20,Laz_Pog22}:
\begin{equation}
    B \approx \sqrt{4\pi \rho\, c_s}\,\frac{M_s}{M_A},
    \label{mm2}
\end{equation}
where $c_s$ is the sound speed of the medium. A key advantage of the MM2 method is that, unlike approaches based on velocity centroids or channel maps, it does not require spectroscopic data, making it applicable even when only synchrotron continuum observations are available.

\section{Discussion}

\subsection{Properties of turbulence reflected by fluctuations of gradient directions}

Super-Alfv\'enic turbulence is widespread in astrophysical environments. In this work, we have demonstrated a new method for characterizing such turbulence using the statistics of magnetic-field directions as inferred from observable gradients \citep{LYP24}. These gradients include velocity gradients \citep{LY18a}, synchrotron intensity gradients \citep{synch_grad17}, and synchrotron polarization gradients \citep{LY18polar}. Here we focus on the statistics of gradient orientations as quantified by the structure function $D^\phi(R)$.

This study complements our earlier work in \citet{dispersion}, where we identified the PDFs of gradient orientations as a useful diagnostic of the Alfv\'en Mach number $M_A$. In contrast, the present analysis shows that $D^\phi$ allows us to recover the scale $l_A$ at which super-Alfvénic turbulence transitions to an MHD regime (see Eq.~\ref{l_A}).

In \citet{LPY25}, we investigated the statistics and spectra of the polarization angles. These measures differ fundamentally from $D^\phi$ and ${\cal F}^\phi$. Although polarization encodes the combined contribution of the regular and turbulent magnetic field ${\bf B} = {\bf B}_{\rm reg} + \delta{\bf B}$, gradient orientations are insensitive to the regular component of the field \citep{LYP24}. This distinction makes $D^\phi$  uniquely sensitive to the transition scale $l_A$, and opens the possibility of combining polarization and gradient diagnostics synergistically.

Obtaining magnetic-field information from gradients requires a sub-block averaging procedure \citep{YL17a}. In practice, this implies that the effective resolution of the gradient maps is reduced by roughly an order of magnitude relative to the original velocity-channel or synchrotron-intensity maps. However, the technique developed in this paper enables the construction of two-dimensional maps of $l_A$, provided that the transition scale is larger than the size of the averaging sub-block.

\subsection{Turbulence in Galaxy clusters and molecular clouds}

Turbulence in galaxy clusters is super-Alfvenic. Polarization was observed in relics \cite{Stuardi21}, which is consistent with the magnetic field maps obtained over larger areas using gradients \cite{Hu_clusters24}. The saturation of polarization direction, polarization degree structure functions (see \cite{LPY25}), or synchrotron intensity structure functions \cite{LP12},  one can obtain $L$, which can be combined with the SIG direction structure functions to find $l_A$. 

Super-Alfv\'enic molecular clouds provide another class of objects where the new approach can be very important. In molecular clouds, there are other ways to establish $L$ from observations, e.g., with spectroscopic measurements (see Eq. (\ref{Dv}). 

\subsection{Obtaining magnetic field strength from observations}
The assumption of equipartition between kinetic and magnetic energies underlies the classical Davis–Chandrasekhar–Fermi (DCF) technique \citep{Davis51,ChanFer53} for estimating the magnetic-field strength. However, this assumption is fundamentally incompatible with the definition of super-Alfv\'enic turbulence, for which $E_k > E_b$ by construction. As a result, on scales larger than $l_A$ the magnetic field is dynamically unimportant: the motions are not constrained by magnetic tension, and therefore the velocity dispersion inferred from Doppler-broadened lines cannot be reliably combined with the dispersion of magnetic field orientations to infer the field strength.

Even in sub-Alfv\'enic turbulence, equipartition is not guaranteed. The applicability of the DCF technique in this regime is further complicated by uncertainties in the driving of turbulence. As demonstrated in \citet{Laz_Y_P25}, velocity-driven sub-Alfv\'enic turbulence can yield a relation of the form
\begin{equation}
    E_m = E_k M_A^2,
\end{equation}
where $E_m$ and $E_k$ denote the magnetic and kinetic energies, respectively. This scaling motivates modified versions of the DCF formula and is consistent with empirical adjustments proposed in earlier works \citep{Bea22,2023A&A...672L...3S}. However, those studies incorrectly attributed the anomalous magnetic-kinetic energy ratio to compressibility effects, whereas it arises from the turbulence-driving mechanism. The crucial difference between sub- and super-Alfv\'enic turbulence is that, in the former, magnetic tension dominates the dynamics on large scales regardless of the driving, making DCF-like modifications feasible. In the super-Alfv\'enic case, no analogous modification is physically justified because magnetic tension does not control motions above $l_A$.

The Differential Measure Approach (DMA) \citep{Laz_Pog22} was proposed as an alternative to DCF. Unlike DCF, DMA does not compare dispersions; instead, it compares the structure functions of velocity and polarization angles with small spatial lags. This enables magnetic-field measurements over localized regions smaller than the turbulence injection scale. However, the DMA requires that the structure functions of velocity and polarization angles have identical slopes—an assumption that, as shown in this paper, is violated in super-Alfv\'enic turbulence.

The method developed here also employs structure functions, but differs fundamentally from the DMA. Rather than relying on small-lag behavior, our technique requires measurements across the range of scales surrounding the transition scale $l_A$. As a consequence, it does not provide fully localized measurements and is therefore conceptually closer to the DCF approach. Nevertheless, it uniquely enables the determination of $l_A$ in super-Alfv\'enic turbulence—a quantity that cannot be obtained with either DCF or DMA.

\section{Summary}

In this paper, we numerically explore the statistical properties of a new observationally accessible measure: the structure function of the positional angles of gradients, $D^\phi(R)$. For sub-Alfv\'enic turbulence, $D^\phi(R)$ exhibits the behavior expected for the structure function of Kolmogorov turbulence: its slope scales as $R^{5/3}$ and saturates at the injection scale of turbulence $L$. 

The most intriguing behavior of $D^\phi(R)$ emerges in the super-Alfv\'enic regime. Because gradient orientations are insensitive to the large-scale mean magnetic field, the structure function $D^\phi(R)$ saturates at the transition scale $l_A$, where super-Alfv\'enic turbulence becomes magnetohydrodynamic. Determining $l_A$ is essential for understanding key astrophysical processes in molecular clouds and galaxy clusters, including heat and cosmic-ray transport, energetic-particle acceleration, and the diffusion of matter and magnetic fields. 

If the turbulence injection scale $L$ is known, the measurement of $l_A$ directly provides the Mach number Alfv\'en $M_A$, a fundamental parameter characterizing magnetized turbulence. We further demonstrate that knowing $l_A$ or $M_A$ allows the magnetic-field strength to be obtained through two independent methods, significantly enhancing the robustness of magnetic-field measurements in super-Alfv\'enic environments.\\

{\bf Acknowledgments}
A.L. acknowledges the support of NSF grants AST 2307840. Y.H. acknowledges the support for this work provided by NASA through the NASA Hubble Fellowship grant \# HST-HF2-51557.001 awarded by the Space Telescope Science Institute, which is operated by the Association of Universities for Research in Astronomy, Incorporated, under NASA contract NAS5-26555. This work used SDSC Expanse CPU, NCSA Delta CPU, and NCSA Delta GPU through allocations PHY230032, PHY230033, PHY230091, PHY230105, PHY230178, and PHY240183 from the Advanced Cyberinfrastructure Coordination Ecosystem: Services \& Support (ACCESS) program, which is supported by National Science Foundation grants \#2138259, \#2138286, \#2138307, \#2137603, and \#2138296.

%\appendix

%\clearpage

%\clearpage
\bibliographystyle{aasjournal}

\bibliography{refs.bib}
\label{lastpage}

% No idea why it goes wrong.
\end{document}